\documentstyle[12pt]{article}

\begin{document}


\hsize=6.15in
\vsize=8.2in
\hoffset=-0.42in
\voffset=-0.3435in

\normalbaselineskip=24pt\normalbaselines

\noindent {\bf BMC Biology 5:18 (2007)}.

\vspace{1.5cm}

\begin{center}
{\large\bf Global and regional brain metabolic scaling and its functional
consequences}
\end{center}

\vspace{0.7cm}

\begin{center}
{\large  Jan Karbowski}
\end{center}

\vspace{0.35cm}


\begin{center}
{\it Sloan-Swartz Center for Theoretical Neurobiology,\\
Division of Biology 216-76, \\
California Institute of Technology, Pasadena, CA 91125, USA }\\
\end{center}

\vspace{0.35cm}

\begin{abstract}
\noindent 
{\bf Background:} Information processing in the brain requires large 
amounts of metabolic energy, the spatial distribution of which is highly 
heterogeneous reflecting complex activity patterns in the mammalian brain. \\
{\bf Results:}
Here, it is found 
based on empirical data that, despite this heterogeneity, the volume-specific 
cerebral glucose metabolic rate of many different brain structures scales 
with brain volume with almost the same exponent around $-0.15$. 
The exception is white matter, the metabolism of which seems to scale with a
standard specific exponent $-1/4$. The scaling exponents for the total 
oxygen and glucose consumptions in the brain in relation to its volume are 
identical and equal to $0.86\pm 0.03$, which is significantly larger than 
the exponents $3/4$ and $2/3$ suggested for whole body basal metabolism on
body mass. \\
{\bf Conclusions:}
These findings show explicitly that in mammals (i) volume-specific scaling 
exponents of the cerebral energy expenditure in different brain parts are 
approximately constant (except brain stem structures),
 and (ii) the total cerebral
metabolic exponent against brain volume is greater than the much-cited
Kleiber's 3/4 exponent. 

The neurophysiological factors that might account for the regional 
uniformity of the exponents and for the excessive scaling of the total 
brain metabolism are discussed, along with the relationship
between brain metabolic scaling and computation.

\end{abstract}

\vspace{0.3cm}


\vspace{0.1cm}


\vspace{0.2cm}

\noindent  email: jkarb@its.caltech.edu



\newpage

\noindent {\bf Background}

\vspace{0.4cm}

The brain is one of the most expensive tissues in the body 
\cite{schmidt,aiello,rolfe}, 
as it uses large amounts of metabolic energy 
for information processing \cite{siesjo,clarke,ames,laughlin}. 
Because of this, neural codes are constrained not only by a combination 
of structural and functional requirements \cite{finlay,barton,clark,
karbowski1,karbowski2,stevens,chklovskii,chklovskii2,sporns,striedter}, 
but also by energetic demands \cite{ames,laughlin,levy1,levy2,attwell,lennie}. 
In general, it is has been observed that an elevated synaptic signaling 
between neurons leads to more energy consumed \cite{sibson}, 
which is used in imaging experiments \cite{vanzetta}. 
Although some theoretical 
progress has been made in quantifying contributions of different 
neurophysiological processes to the total metabolic expenditure of 
a single neuron \cite{attwell}, 
the metabolism of large-scale
neural circuits has not been investigated quantitatively. Here, a first 
step in this direction is made by studying global and regional {\it in vivo} 
brain metabolic scaling. There is a long tradition in applying allometric 
scaling to problems in biology \cite{schmidt}, in particular to 
whole body metabolism \cite{schmidt,kleiber}, 
but surprisingly not to cerebral metabolism.  
The goal of this paper is to find, by collecting and analyzing data, 
scaling metabolic exponents of different parts of the brain, as
well as its global exponent. It is found that the volume-specific exponents
across cerebral regions on brain volume are almost identical, 
approximately $-0.15$. 
Consequently, the energy consumption of the entire brain tissue scales
with brain volume with the exponent $\approx 0.86$. The main
neurophysiological factors that might cause the increase of the latter 
exponent well above the putative Kleiber's 3/4 scaling exponent characterizing
whole-body energy expenditure on body mass \cite{kleiber}, are identified. 
The consequences of brain metabolic scaling on its information processing 
capacity in the context of brain design are also discussed.


\vspace{1cm}

\noindent {\bf Results}

\vspace{0.4cm}

Oxygen and glucose are the main components involved in the production of 
ATP, which is used in cellular energetics \cite{rolfe,siesjo,ames}, and 
therefore their rates of utilization provide useful measures of brain 
metabolism. There are several mammalian species, spanning more than 3 orders 
of magnitude in brain size, for which {\it in vivo} brain metabolic data 
are available (see Materials and Methods). The allometric laws 
characterizing global cerebral metabolism of oxygen and glucose are
similar and yield an identical scaling exponent $0.86\pm 0.03$ against
brain volume (Fig. 1). 
It is important to note that this value is significantly larger 
($p \le 0.05$) than the exponents 3/4 \cite{schmidt,kleiber} 
and 2/3 \cite{white} found for whole body mammalian metabolism in relation
to body mass.

The cerebral cortex is a critical part of the brain responsible for 
integrating sensory information, and commanding behavioral and cognitive tasks.
Regions of the cerebral cortex differ both in molecular detail
and in biological function, which is manifested in a non-uniform distribution 
of neuronal activity and energy utilization throughout the cortex 
\cite{clarke} (and Supplementary Information). 
However, despite this heterogeneity values of the scaling 
exponents of the regional volume-specific glucose utilization rates on
brain volume (CMR$_{glc}$; glucose cerebral metabolic rate per brain region
volume) are surprisingly homogeneous; they are either exactly equal to 
$-0.15$ or are close to this value (Fig. 2 and Table 1). 
Consequently, also the 
average specific glucose utilization rate of the whole cerebral cortex 
scales with the exponent $-0.15\pm 0.03$ with brain volume (Fig. 2E), 
which is equivalent to the exponent $0.85\pm 0.03$ for the metabolism of 
the entire cortical volume; the value close to that for the whole brain 
(Fig. 1).

Some subcortical structures of gray matter utilize half of the energy
used in the cortex (e.g., limbic structures in cat and monkey; see Clarke
and Sokoloff \cite{clarke} and Supplementary Information), 
and yet almost all of them exhibit a similar scaling homogeneity, with 
metabolic specific exponents also around $-0.15$ (Fig. 3, and Table 1). 
Volume-specific metabolisms of two brain stem 
structures - superior colliculus (involved in visual coordination) and 
inferior colliculus (involved in auditory processing) - seem to be 
exceptions, since they scale with brain size with the exponent $\approx 0$ 
(Table 1). This might be caused by their highly variable
activities (see Supplementary Information), and we do not know how other 
brain stem structures behave.
The high degree of allometric uniformity for the most 
of the subcortical system is even more striking than that for the cortex, 
since subcortical regions are much more diverse in function and biophysical 
properties than cortical areas. 
For example, thalamus and hippocampus play extremely different roles 
in the brain, the former mediating sensory input to the cortex, while the 
latter implicated in memory processes, and still their scaling exponents
and corresponding confidence intervals are almost identical (Figs. 3A,B, 
and Table 1).

Metabolism of white matter is about three-fold lower than that of 
gray matter \cite{siesjo,clarke}, and the results in 
Table 1 indicate that also their scaling exponents might differ.
Specific glucose metabolisms of the two main structures of white matter, 
corpus callosum and internal capsule, scale similarly with brain volume 
with the average exponent around $-1/4$ (Fig. 4, Table 1). This value
resembles the whole body specific metabolic exponent against body mass.


\vspace{1.5cm}

\noindent {\bf Discussion}

\vspace{0.4cm}

\noindent {\it General discussion.}

\vspace{0.4cm}

The total metabolic exponent $0.86\pm 0.03$, found for unanesthetized 
mammalian brains, implies that cerebral energy use increases more steeply 
with brain size 
than does whole body energy use with body size. This might be a reason why 
brain size increases 
slower than its body size, with the power 3/4 \cite{martin1}. 
However, the metabolic exponent 0.86 undermines Martin's \cite{martin1,martin2}
well-known argument that the 3/4 scaling of brain size reflects 
allometric isometry between brain metabolic needs and its size (exponent 1). 
Instead, the product of the exponents for brain size on body size and 
brain metabolic rate on brain size gives an exponent for brain metabolic 
needs on body size of about 0.65, which is lower than 0.75, expected
from the assumption of isometry. Thus, depending on the scaling reference
the total brain metabolism scales either above the Kleiber's 3/4 power
if in relation to brain mass, or below the Kleiber's power if in relation
to body mass.

The discovered uniformity, i.e. constancy, of the cerebral specific exponents 
in gray matter suggests a common principle underlying basal metabolism of 
different brain structures, which might be associated with the homogeneity of 
synaptic density throughout the gray matter \cite{braitenberg,cragg};
see below.

In analyzing comparative allometry some authors use phylogenetic approaches
in order to include dependencies in data sets \cite{harvey,garland}. 
However, these sophisticated methods require as a prerequisite the knowledge
of a phylogenetic tree and associated branching parameters for species
of interest. This is not a trivial matter to do, and therefore not applied
in the present analysis. Also, because correlations in most scaling plots
(Figs. 1-4, Table 1) are generally very high, it is likely that taking
phylogeny into account would not alter the empirical exponents.

\vspace{0.4cm}

\noindent {\it Key factors in the cerebral metabolic rate.}

\vspace{0.4cm}

Which factors in the gray matter might account for its total metabolic 
exponent on brain volume greater than 3/4? The likely candidate is the density
of glial cells, which provide metabolic support for neurons (including synapses)
\cite{tsacopoulos,magistretti}. Two recent studies \cite{herculano,sherwood}
show that the number of glia per neuron increases for larger brains,
suggesting that neurons become more energy expensive with increasing brain
size. In particular, the total number of glia in the cerebral cortex of
rodents scales with brain size with the exponent 0.89 
(Fig. 2B in \cite{herculano}),
which is close to the empirical metabolic exponent 0.86. 
It is likely that glia number simply follows energetic demands of neurons,
especially their spatial expansion with increasing brain size. Thus,
the details of this expansion may provide clue about which neural factor
are important.

It has been estimated that neural metabolism is dominated by 
Na$^{+}$/K$^{+}$-ATPase ion transport \cite{ames,astrup,erecinska,hulbert}. 
The bulk of this comes from active ion fluxes at synapses and along axon
(propagation of axon potentials) if neural firing rate is 
sufficiently large \cite{attwell}. The synaptic contribution in a single neuron
is proportional to the product of the number of synapses per neuron, 
firing rate, release probability, and the postsynaptic charge. 
The active axon contribution is proportional to its surface area and 
firing rate. When neurons do not fire action potentials they also consume 
energy, because of the passive Na$^{+}$ and K$^{+}$ ion flow that 
electro-chemical Na$^{+}$/K$^{+}$ pump must remove to maintain their 
gradients across the membrane \cite{kandel}. This resting potential 
contribution is proportional to the total neuron's surface area. 
To obtain the total cerebral energy consumption one has to multiply all
these three additive contributions by the total number of neurons in the 
gray matter.

It seems that from all these three neural contributions the most dramatic for 
the scaling exponent of the total brain metabolism on brain volume is the 
synaptic contribution. This contribution is proportional to the total number 
of synapses in the gray matter, which scales with the gray matter volume, 
and thus brain volume \cite{prothero,hofman}, with the exponent 1.
This follows from regional- and scale-invariance of synaptic density
\cite{braitenberg,cragg}. The regional homogeneity of synaptic density
correlates with the discovered homogeneity of the regional volume-specific 
cerebral metabolic scaling (Table 1). Moreover, if remaining factors comprising
the synaptic contribution, i.e., firing rate, release probability, and 
postsynaptic charge were brain size independent, then the synaptically driven 
total brain metabolism would scale with brain volume with the exponent 1. 
Since this exponent is, in fact, between 3/4 and 1 (Fig. 1), 
it implies that either of
these 3 factors (or all of them) decrease with brain size. It could be
hypothesized that because synaptic sizes and their basic molecular 
machinery are similar among mammals of different sizes 
\cite{braitenberg,kandel}, the 
postsynaptic component might be roughly the same among different species.
This suggests that the product of the firing rate and release probability
should decrease as brains increase in size, with a power of about $-0.15$,
which is in accord with low firing rates in humans estimated based 
on their basal cerebral metabolic rate \cite{lennie,shoham}. 
This also suggests that the number of active synapses in 
the background state decreases for bigger brains. That firing rate should
decrease with brain (body) size is also consistent with allometric data
of avian sensory neurons firing rates \cite{hempleman}.

The remaining two contributions affecting metabolic rate, the active axons
and maintaining the resting potentials, are both proportional to the 
product of the number of neurons in the gray matter and axonal surface area
(assuming that axon surface area constitutes the majority of the neuron's area,
especially for bigger brains). Additionally, the active axon contribution
is proportional to the average firing rate. Given the above indications
that the firing rate likely decreases with brain volume the active axon
contribution becomes less important for the total metabolic exponent as brains
increase in size. Thus, we focus only on the resting potential contribution.
Since the product of the number of neurons and axonal surface area
is proportional to the ratio of the volume of gray matter to axon diameter
(due to the empirical fact that volumes of intracortical axons and gray matter
are proportional across species \cite{braitenberg}), the resting potential
contribution produces the total metabolic exponent that is determined by the 
scaling exponent of the axon diameter against brain volume.
The bigger the axonal exponent, the smaller the 
metabolic exponent. There are some sketchy experimental indications
that axon diameter indeed increases with brain size \cite{olivares,harrison}, 
e.g., in the corpus callosum with the exponent
$\approx 0.07$ \cite{olivares}. If similar allometry holds for
the gray matter axons, the resting potential contribution would yield
a total metabolic exponent also above 3/4.

\vspace{0.4cm}

\noindent {\it Brain metabolism, computation, and design.}

\vspace{0.4cm}

The facts that the synaptic metabolic contribution decreases with 
decreasing the rate of release probability, and that the active axon and
resting potential metabolic contributions decrease with increasing axon 
diameter have interesting functional consequences. Higher failure rate of synaptic 
transmission as brains get bigger not only saves energy but it also
may maximize information transfer to postsynaptic neurons 
\cite{levy2}. Similarly, increasing axon diameter with brain
size accomplishes three functions simultaneously: it reduces the specific
metabolic rate, increases the number of synapses per neuron, and it 
increases the speed of signal propagation in cortical circuits, which is 
proportional to the square root of the axon diameter 
\cite{hodgkin,koch}. It is difficult to say at this stage whether these
relationships are accidental or maybe a result of some cerebral optimization.

We can estimate the allometric cost of information processing by finding 
how the amounts of metabolic energy per neuron and per synapse scale 
with brain size. Since the total energy utilized by the entire cerebral
cortex (gray matter) scales with its volume $V_{g}$ with the exponent
$\approx 0.85$ (Table 1, Fig. 2E), the cerebral energy per neuron is
$\propto V_{g}^{0.85}/(\rho_{n}V_{g}) \propto 
V_{g}^{0.05-0.17}$, i.e., it increases with brain size, where we used
the fact that the neural density $\rho_{n}$ scales with brain volume with
the exponent between $-0.32$ and $-0.20$ (based on data of Haug (1987)
\cite{haug}; see Supplementary Fig. S2). This increase is in accord with
the trend of increasing the number of glial cells per neuron in gray matter.
The opposite is true for synapses, since the energy per synapse is
$\propto V_{g}^{0.85}/(\rho_{s}V_{g}) \propto V_{g}^{-0.15}$, 
where $\rho_{s}$ is the synaptic density (independent of
brain size). The increase in energy expenditure per neuron with increasing
brain size is in contrast to findings in liver cells \cite{porter},
whose metabolism decreases with body mass. This difference
reflects the increase of neural size (its wire) and corresponding decrease 
in density with increasing brain volume (sizes of liver cells
are roughly constant \cite{purves}). The decaying trend for synapses implies 
again that their 
active fraction decreases as brains get bigger. These results have 
implications for coding and cortical organization. The fact that expanding 
neurons are energetically costly was probably a driving evolutionary force in 
decreasing their density in larger mammals (Supplementary Fig. S2) 
and adopting sparse neural representations \cite{levy1,attwell,lennie}.
The latter factor is consistent with the idea of functional modularity
of the cerebral cortex \cite{prothero}, i.e., that primary information
processing takes place in local modules/areas. The sizes of such modules/areas
seem to follow scaling rules \cite{karbowski2,chklovskii2,prothero}, and
have been shown to have almost brain size independent connectedness 
\cite{karbowski2} as opposed to neural connectedness that decays with brain 
size \cite{karbowski1,karbowski2}.

\vspace{0.4cm}

\noindent {\it Metabolism of gray versus white matter.}

\vspace{0.4cm}

The data in Table 1 seem to indicate that the white matter and gray matter 
metabolic allometries are different. This finding implies that as brains 
increase in size, the white matter metabolism is less costly than that of
gray matter. This is presumably beneficial for the total 
cerebral energetic expenditure, since white matter increases 
disproportionately faster than gray matter \cite{barton,prothero,zhang}. 
The difference in white and gray matter metabolisms may be caused by 
their apparent neuroanatomical differences, since most of the white matter 
axonal membrane is covered with myelin sheath, which prevents ions flow
and reduces metabolic cost.


\vspace{0.4cm}

\noindent {\it Relation to metabolism of other tissues.}

\vspace{0.4cm}

Most of the tissues in the body have much lower specific metabolic rate 
than the brain, with the exception of four highly active organs: 
kidney, liver, heart \cite{schmidt} and gut \cite{aiello}. 
There exists no reliable 
data on {\it in vivo} allometric metabolic scaling in these tissues across
different species (see however \cite{wang}, where allometric exponents
are given based on only 2-3 species). Allometric {\it in vitro} studies
of Na$^{+}$/K$^{+}$-ATPase in kidney \cite{turner1} and in brain 
\cite{turner2} suggest that these two organs might have comparable 
specific metabolic exponents. We can indirectly estimate and compare 
metabolic exponents for active tissues using allometric data on 
mitochondria size \cite{else}, since its total membrane 
surface area correlates with a baseline oxygen consumption 
\cite{else}. (Interestingly, the total mitochondrial volume in locomotory
muscles is proportional to the maximal metabolic rate in mammals 
\cite{weibel}.) We find that the total 
mitochondrial surface area in brain scales with brain mass with the 
exponent 0.86, i.e., exactly the same as that in Fig. 1. 
The corresponding scaling exponents for kidney, liver,
and heart, against their respective organ masses are smaller and closer
to the 3/4 exponent: 0.71, 0.74, and 0.81. These exponents do not
seem to relate directly to the exponents of sizes of these tissues 
on body size, since masses of kidney and liver increase slower than body 
mass, while heart mass scales isometrically \cite{schmidt}.
Thus, it appears that, in general, higher metabolic exponents do not 
necessarily lead to lower mass exponents.

If these exponents reflect a real difference of metabolic allometry 
between cerebral and non-cerebral tissues, it might be caused by differences
in membrane chemical composition and ion pump activities, which is known as
the ``membrane pacemaker theory of metabolism'' \cite{hulbert}.
For example, it has been shown that heart, liver, kidney, and skeletal 
muscles display allometric variation in lipid composition but brain does 
not \cite{hulbert2}. Other potential factors (some of which might be 
related to the membrane pacemaker
hypothesis) affecting differences in the allometries of brain and
other tissues include:
(i) distinct ways energy is utilized in the brain and in other tissues
(e.g., Na$^{+}$/K$^{+}$-ATPase dominates energy consumption only in the 
brain and kidney \cite{rolfe}); (ii) difference in a mode of 
activity (cells outside nervous system exhibit graded electrical activity 
without firing Na$^{+}$ action potentials); 
(iii) structural differences between brain and other highly active 
tissues (the size of non-cerebral cells is virtually independent of 
the body mass and the cells lack elongated processes with synapses, 
e.g., \cite{purves}); (iv) the existence of the blood-brain barrier 
that restricts a direct transport of molecules between bloodstream and 
nerve cells \cite{kandel}, which might affect substrate utilization rate
and/or neural activity \cite{devor,sheth}.

\vspace{0.4cm}

\noindent {\it Supply-limited models of metabolic scaling.}

\vspace{0.4cm}

In many studies of whole body metabolism the scaling exponent 3/4 was 
found \cite{schmidt,kleiber}, and it was argued that 
this value follows from a general model of hierarchical fractal-like 
transport networks \cite{west}, or from constrained geometric 
networks with balanced supply and demand \cite{banavar}. Both of these models
are based on the assumption that metabolic rates are determined solely by
resources supply rates and are independent of the cellular energy expenditure.
This single-cause assumption has been challenged recently 
\cite{darveau,suarez}.
The main arguments against the above supply-limited models are that (i) blood
flow rate adjusts itself to tissue physiological demands and in resting animals
is well below its maximal limit, and (ii) cellular metabolic rates decline
with increasing body size \cite{porter} (at least for non-cerebral cells). 
Because of these facts a ``multiple-causes'' scenario of metabolic scaling 
has been proposed \cite{darveau,suarez}, which in essence argues that 
supply rate is only one of the factors and should be considered together 
with other factors characterizing utilization of cellular energy.
The approach taken in this paper is similar in spirit, i.e., given that 
the total cerebral metabolic exponent is 0.86, simple supply-limited models 
are rejected as a possible explanation for brain metabolic scaling. 
Instead, the most energetically expensive cellular factors that are most 
likely to affect the metabolic exponent were identified. 
In this sense, this approach can also be viewed as 
a multiple-cause model of the cerebral metabolic scaling.

\vspace{1.5cm}

\noindent {\bf Conclusions}

\vspace{0.4cm}

Figures 1-3 and Table 1 provide an empirical evidence that the scaling 
exponents describing global and regional brain metabolism are significantly 
different ($p \le 0.05$) from the much-cited 3/4 exponent, which calls into 
question the direct applicability of supply-limited models \cite{west,banavar} 
to brain metabolism. The exceptions are white matter structures, which seem to 
exhibit ``regular'' metabolic exponents (Fig. 4).

The empirical results presented in this paper show striking uniformity of 
the allometric metabolic exponents over almost entire gray matter of
mammalian brains at normal resting conditions, despite anatomical and
functional heterogeneity of different regions and their different levels
of activation. This regional scaling uniformity is surprising, as activity
level could potentially affect the scaling exponent. For example, the total 
metabolic rate of maximally exercised body scales with body mass well above
3/4, with an exponent of 0.88 \cite{weibel,bishop}.


\vspace{1.4cm}

\noindent {\bf Materials and Methods}

\vspace{0.4cm}

{\it In vivo} data of the cerebral oxygen (CMR$_{O_{2}}$) and glucose 
(CMR$_{glc}$) utilization rates of unanesthetized adult animals at 
resting conditions were collected from different sources \cite{data}
(see Supplementary Information for details). 
In those studies the measurements of glucose utilization in all species were 
performed by essentially the same method or its modification 
(in human and baboon),
and thus all glucose data are directly comparable. 
There is a small method variability for oxygen data, since the same technique 
was applied to five out of seven species (except cat and dog). However, this
variability does not affect the scaling exponent (Fig. 1A) - it is exactly
the same even if only single-method mammals are included in the plot.
Glucose utilization data represent both global and regional cerebral 
metabolism. The investigated mammals include: Swiss mouse (only glucose data), 
Sprague-Dawley rat, squirrel (only glucose data), rabbit (only glucose data), 
cat, dog (only oxygen data), macaque monkey, baboon, sheep, goat 
(only glucose data), and human. The investigated brain structures include: 
cerebral cortex (visual, prefrontal, frontal, sensorimotor, parietal, 
temporal, cingulate, occipital), thalamus (including lateral geniculate
nucleus and medial geniculate nucleus), hypothalamus 
(and separately mammillary body), cerebellum (including cerebellar cortex 
and dentate nucleus), basal ganglia (caudate, substantia nigra, 
globus pallidus), limbic system (hippocampus, amygdala, septum), 
brain stem (superior colliculus, inferior colliculus), and white matter 
(corpus callosum, internal capsule). In the cases when there are more than 
one data point for a given animal or a brain structure, an arithmetic mean 
of all values was taken.

Allometric metabolism of the entire cerebral cortex, presented 
in Fig. 2E, was obtained by computing an arithmetic mean of glucose 
utilization in 8 cortical areas (listed above) for each animal.
Glucose utilization of a given area was itself an average of values
taken from different sources. If data for all 8 areas were not available, 
averaging was performed over lesser number of areas. For consistency, 
also an alternative method of averaging was used: first averaging was
performed for a given source data, and second among different 
sources representing the same animal. In this method, because various 
sources differ in the number of cortical areas studied, averaging
in many cases was performed over significantly different number of areas.
However, both methods give statistically identical scaling exponents for 
the cerebral cortex metabolism (see Fig. S1 in Supplementary Information).

Allometric metabolism of the entire brain was obtained by using 
either direct data quoted by authors, or, if not available, by computing 
an arithmetic mean of glucose consumption in all brain structures provided 
by the authors.
For all scaling plots brain volumes were taken from Hofman (1988) 
\cite{hofman} and Stephan et al (1981) \cite{stephan}, or from the source.

\vspace{1.2cm}

\noindent{\bf Acknowledgments}

\noindent
The work was supported by the Sloan-Swartz Fellowship and by the Caltech Center
for Biological Circuit Design.

\vspace{1.5cm}


\newpage

\begin{center}

\begin{tabular}{|l c c c c|}
\multicolumn{5}{l}
{ Table 1.}\\
\multicolumn{5}{l} 
{Specific scaling exponents of the regional cerebral glucose utilization 
rate CMR$_{glc}$} \\
\multicolumn{5}{l} 
{against brain volume. } \\
  \hline \hline

 Brain    &  Scaling   &  95$\%$ confidence  &   Correlation  &  Number of \\
structure &  exponent  &   intervals    &  $R^{2}$ ($p$-value) &  species   \\
\hline\hline

{\it Cerebral cortex} & $-0.15\pm 0.03$ & (-0.22,-0.08) & 0.870 (0.0022) & 7\\

 $\;\;$  Visual  &  $-0.12\pm 0.03$ & (-0.17,-0.08) &  0.932 (0.0018)  &  6 \\

 $\;\;$  Prefrontal & $-0.17\pm 0.03$ & (-0.23,-0.10) & 0.953 (0.0044)  & 5 \\

 $\;\;$  Frontal & $-0.14\pm 0.02$ & (-0.17,-0.12)  &  0.997 (0.0015)  &  4 \\

 $\;\;$  Sensorimotor & $-0.15\pm 0.02$ & (-0.19,-0.11) & 0.945 (0.0002) & 7 \\

 $\;\;$  Parietal & $-0.15\pm 0.03$ & (-0.20,-0.11) & 0.954 (0.0008) & 6  \\

 $\;\;$  Temporal & $-0.15\pm 0.05$ & (-0.27,-0.03) & 0.680 (0.0224)  & 7 \\

 $\;\;$  Cingulate & $-0.16\pm 0.03$ & (-0.23,-0.09) & 0.912 (0.0030) & 6  \\

 $\;\;$  Occipital & $-0.20\pm 0.12$ & (-0.53,0.12) & 0.563 (0.1439) &  5  \\

{\it Thalamus}   & $-0.15\pm 0.03$ & (-0.22,-0.08) & 0.858 (0.0027) &  7  \\

{\it Hypothalamus}  & $-0.10\pm 0.04$ & (-0.20,-0.01) & 0.692 (0.0402) &  6 \\

 $\;\;$ Mammillary body & $-0.15\pm 0.07$ & (-0.30,0.00) & 0.773 (0.0495) & 5 \\

{\it Cerebellum}  & $-0.15\pm 0.04$ & (-0.24,-0.06) & 0.840 (0.0102) &  6  \\

{\it Basal ganglia}    &    &    &   & \\

  $\;\;$ Caudate  & $-0.15\pm 0.03$ & (-0.22,-0.08)  & 0.905 (0.0035) &  6  \\

  $\;\;$ Substantia nigra & $-0.14\pm 0.02$ & (-0.18,-0.10) & 0.956 (0.0007)  &  6 \\

  $\;\;$ Globus pallidus  & $-0.16\pm 0.04$ & (-0.24,-0.08) & 0.926 (0.0088)  &  5  \\

{\it Limbic system}    &    &    &   & \\

  $\;\;$ Hippocampus  & $-0.14\pm 0.03$ & (-0.19,-0.09) & 0.919 (0.0007) & 7 \\

  $\;\;$ Amygdala  & $-0.12\pm 0.03$ & (-0.17,-0.07) & 0.919 (0.0025) & 6 \\

  $\;\;$ Septum  & $-0.16\pm 0.03$  & (-0.23,-0.09) &  0.944 (0.0057) & 5  \\

{\it Brain stem}    &    &    &   & \\

  $\;\;$ Superior colliculus & $-0.06\pm 0.05$ & (-0.19,0.06) & 0.448 (0.2168)   &  5  \\

  $\;\;$ Inferior colliculus & $0.05\pm 0.09$ & (-0.14,0.23) & 0.174 (0.4843) &  5  \\

{\it White matter}    &    &    &   & \\

  $\;\;$ Corpus callosum & $-0.23\pm 0.09$ & (-0.40,-0.06) & 0.947 (0.0271) & 4  \\

  $\;\;$ Internal capsule & $-0.24\pm 0.12$ & (-0.48,0.00) & 0.902 (0.0504) & 4  \\

\hline \hline

\end{tabular}

\end{center}

\noindent  t-test shows that population mean of the exponents in the second
column is significantly greater than the exponent $-0.25$  
($p= 9\cdot 10^{-9}$, df=21 if white matter included;
$p= 2\cdot 10^{-9}$, df=19 if white matter excluded).

\newpage

{\bf \large Figure Legends}

\vspace{0.2cm}

Fig. 1\\
Scaling of the total basal cerebral metabolism with brain volume. 
The least-square fit line for the log-log plot yields:
(A) For the total oxygen consumption rate the scaling
exponent is $0.86\pm 0.04$ 
($y=0.86x-1.02$, $R^{2}= 0.989$, $p < 10^{-4}$, $N= 7$), 
and its 95$\%$ confidence interval is $(0.75, 0.96)$. (B) For the total 
glucose utilization rate an identical exponent $0.86\pm 0.03$ is found
($y=0.86x-0.09$, $R^{2}= 0.994$, $p < 10^{-4}$, $N= 10$), 
and its 95$\%$ confidence interval is $(0.80, 0.91)$.

\vspace{0.2cm}

Fig. 2\\
Scaling of the cerebral cortex specific glucose utilization rate, 
CMR$_{glc}$, with brain volume. The specific metabolic scaling exponent, 
corresponding to the slope of the regression line, has the following values: 
(A) $-0.12$ for visual cortex ($y=-0.12x+0.02$); 
(B) $-0.15$ for parietal cortex ($y=-0.15x+0.01$);
(C) $-0.15$ for sensorimotor cortex ($y=-0.15x+0.02$); 
(D) $-0.15$ for temporal cortex ($y=-0.15x+0.07$).
(E) Average glucose utilization rate of the entire cerebral 
cortex yields the specific exponent $-0.15$ ($y=-0.15x+0.03$).

\vspace{0.2cm}

Fig. 3\\
Scaling of the specific glucose utilization rate in subcortical gray matter 
with brain volume. The specific metabolic scaling exponent 
has the following values: 
(A) $-0.15$ for thalamus ($y=-0.15x+0.03$);
(B) $-0.14$ for hippocampus ($y=-0.14x-0.13$), 
which represents limbic structures;
(C) $-0.15$ for caudate ($y=-0.15x+0.02$), which represents basal ganglia;
(D) $-0.15$ for cerebellum ($y=-0.15x-0.09$).

\vspace{0.2cm}

Fig. 4\\
Scaling of the volume-specific glucose
utilization rate in white matter with brain volume. (A) Corpus 
callosum metabolism yields the exponent $-0.23$ ($y= -0.23x-0.45$),
and (B) internal capsule has a similar exponent $-0.24$ ($y= -0.24x-0.41$).

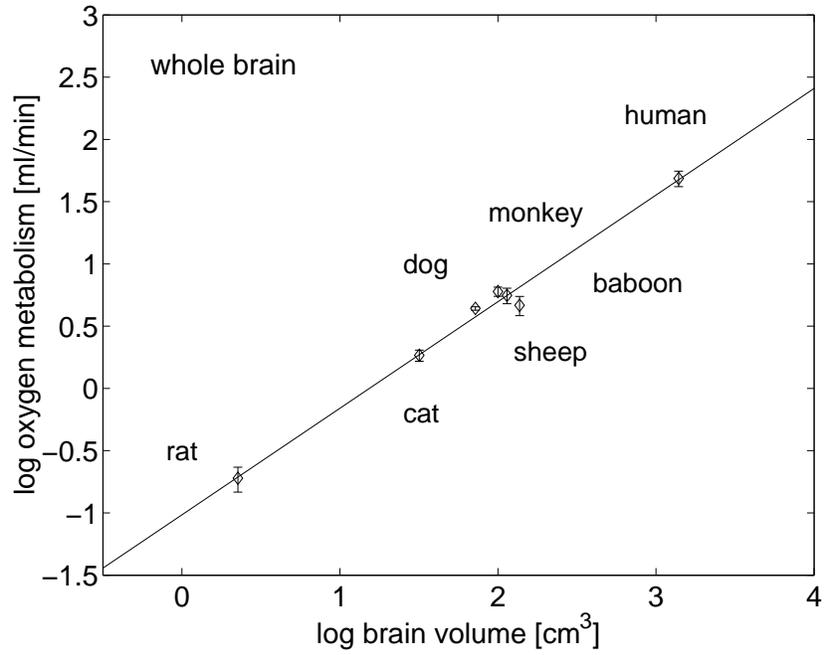
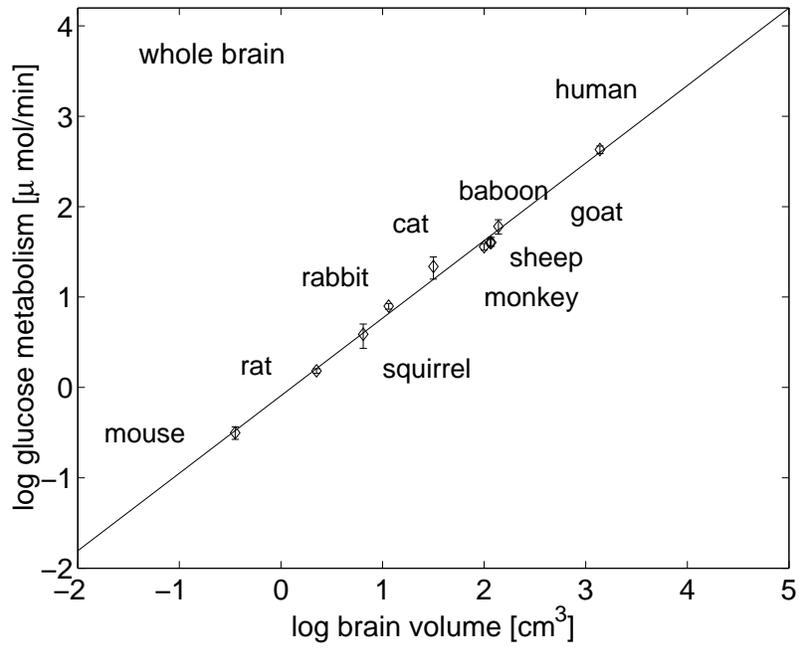
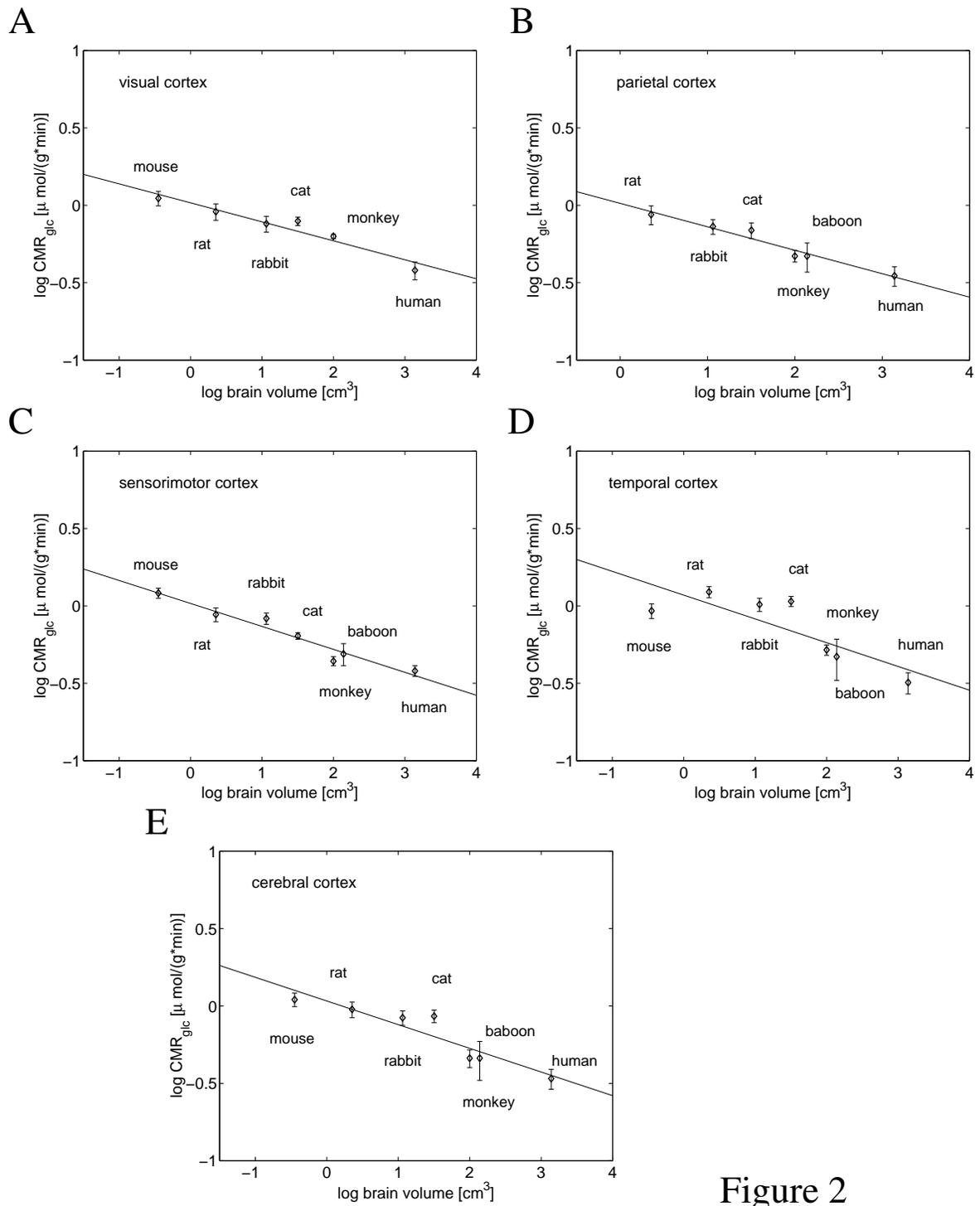
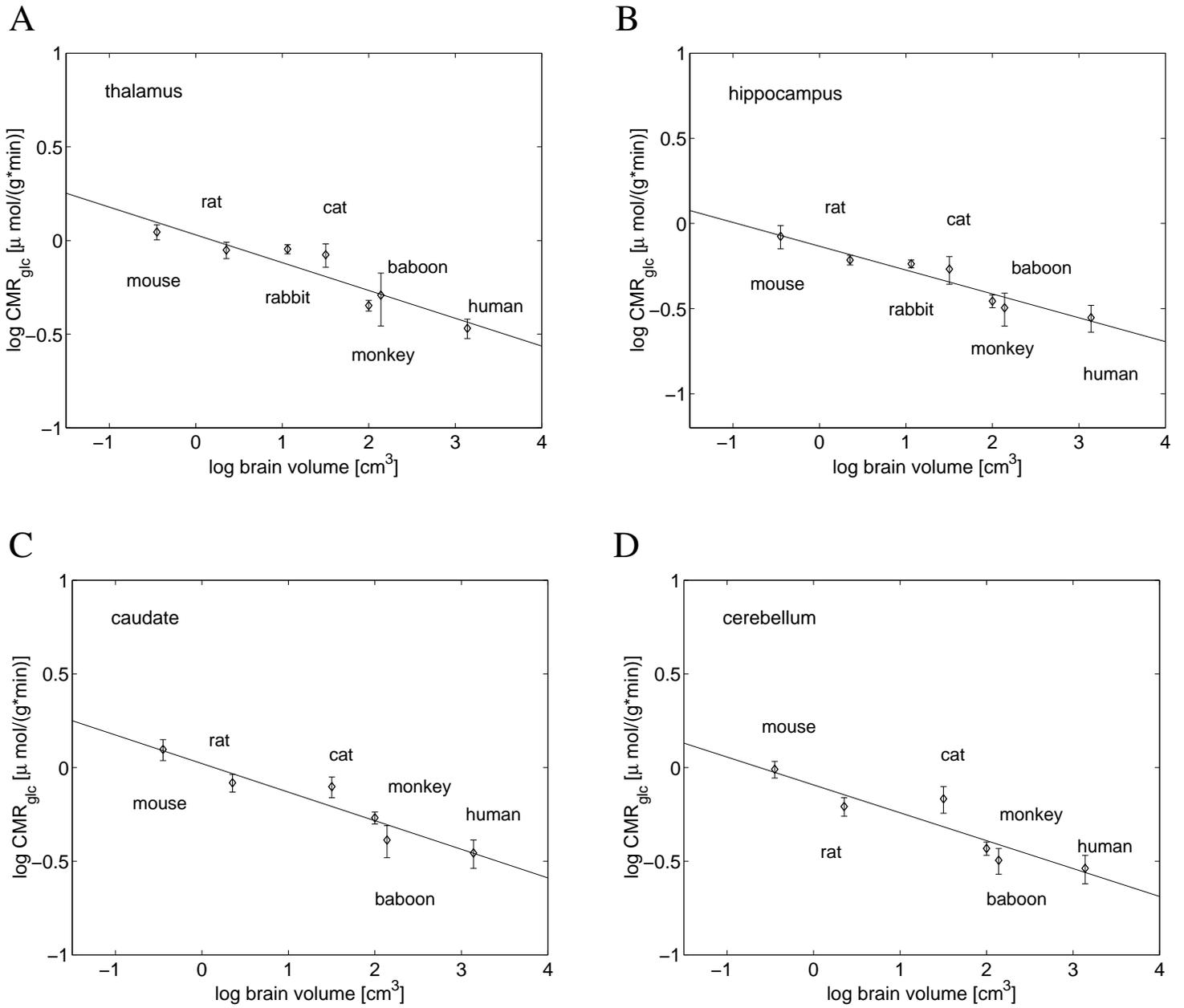
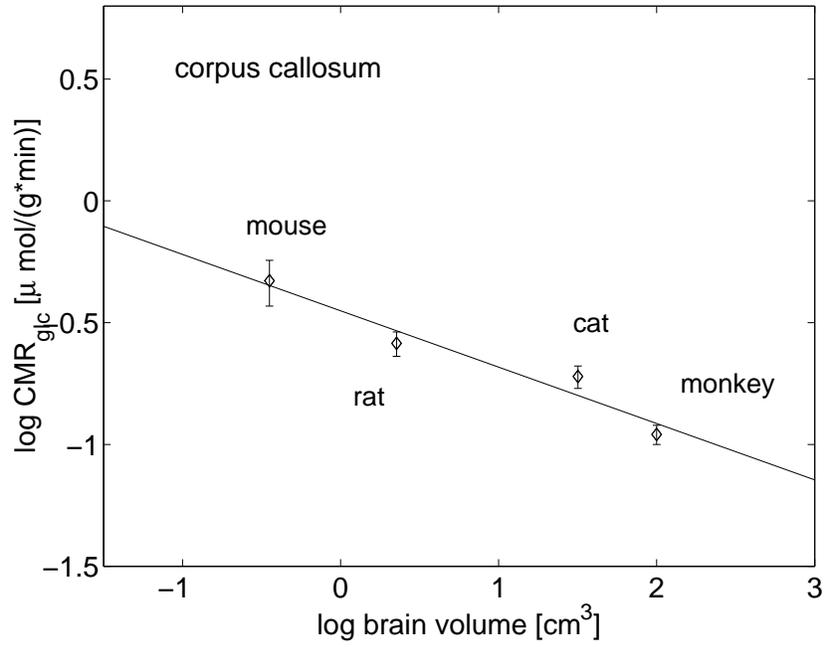
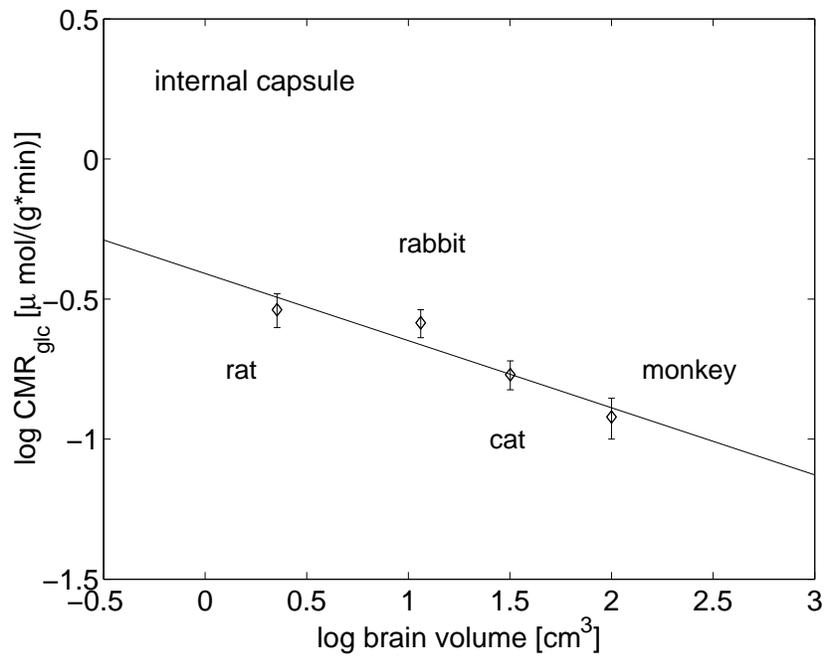
\end{document}